\newcommand{\PRL}[3]{Phys.\ Rev.\ Lett.\ {\bf #1},\ #2 (#3)}

\newcommand{\SC}[3]{Science\ {\bf #1},\ #2 (#3)}

\newcommand{\PRB}[3]{Phys.\ Rev.\ B\ {\bf #1},\ #2 (#3)}

\newcommand{\appsection}[1]{\let\oldthesection\thesection
  \renewcommand{\thesection}{ \oldthesection}
  \section{#1}\let\thesection\oldthesection}

\documentclass[twocolumn,showpacs,preprintnumbers,amsmath,amsfonts,amssymb,notitlepage,pra]{revtex4-1}
\usepackage{geometry}
\date{\today}
\DeclareMathAlphabet{\mathpzc}{OT1}{pzc}{m}{it}
\usepackage[utf8x]{inputenc}
\usepackage{amsmath}
\usepackage{amsfonts}
\usepackage{amssymb}
\usepackage{graphicx}
\usepackage{nicefrac}
\usepackage{amsbsy}
\usepackage{float}
\usepackage{epstopdf}
\usepackage{dsfont}
\usepackage{siunitx}

\def \be{\begin{equation}}
\def \ee{\end{equation}}
\def \ba{\begin{array}}
\def \ea{\end{array}}
\def \bea{\begin{eqnarray}}
\def \eea{\end{eqnarray}}

\begin{document}
\title{Stability of a Floquet Bose-Einstein condensate in a one-dimensional optical lattice}
\author{Sayan Choudhury}
\email{sc2385@cornell.edu}
\author{Erich J Mueller}
\email{em256@cornell.edu}
\affiliation{Laboratory of Atomic and Solid State Physics, Cornell University, Ithaca, New York}
\begin{abstract}
Motivated by recent experimental observations (C.V. Parker {\it et al.}, Nature Physics, {\bf 9}, 769 (2013)), we analyze the stability of a Bose-Einstein condensate (BEC) in a one-dimensional lattice subjected to periodic shaking. In such a system there is no thermodynamic ground state, but there may be a long-lived steady-state, described as an eigenstate of a ``Floquet Hamiltonian". We calculate how scattering processes lead to a decay of the Floquet state. We map out the phase diagram of the system and find regions where the BEC is stable and regions where the BEC is unstable against atomic collisions. We show that Parker et al. perform their experiment in the stable region, which accounts for the long life-time of the condensate ($\sim$ 1 second). We also estimate the scattering rate of the bosons in the region where the BEC is unstable.
 \end{abstract}

\maketitle

\section{Introduction}
Recently there has been much interest in periodically driven quantum systems (Floquet systems), as time dependent forces provide a new knob for accessing interesting phenomena. Some of these phenomena are analogous to physics seen in static systems (e.g edge modes in Floquet topological insulators and artificial gauge fields in cold atom systems) \cite{MoessnerFTIReview,HolthausFloquetReview,FerrariACTunnelingNatPhys2009,SengstockFrustratedScience2011,SengstockIsing2013,SengstockAbelianGaugePRL2012,SengstockNonAbelianGaugePRL2012,GalitskiFTINatPhys2011,RechtsmanPFTI,ZhaiFTIarxiv,GaliskiFTIPRB2013,GedikFloquetScience,GalitskiFregosoFTIPRB2013,Gomez-LeonArxivAQHE,PodolskyFTI2013,BarangerFloquetMajorana,DemlerMajoranaPRL2011,DemlerFloquetTransport, KunduSeradjehMajoranaPRL, OhMajoranPRB2013, DuttaSenPRB2013, BaurCooperFloquet2014,NeupertFloquetPRL2014,TorresPRBGraphene2012,TorresPRBGraphene2014}, but other phenomena are unique to the non-equilibrium system (such as ac-induced tunneling and anamalous edge states in insulators with zero Chern number) \cite{EckardtACEPL,DemlerFloquetAnamolous,LevinFloquetAnamolous,ZhaoPRLAnamolous,MuellerFloquetAnamolous}. In the cold atom context, particular interest has focussed on bosonic systems, as they are most accessible experimentally. Parker et al. recently observed an interesting analog of a ferromagnetic transition in a Bose gas trapped in a shaken one dimensional optical lattice \cite{ChinFloquet2013}. Here, we theoretically analyze their experiment, studying the stability of their condensate. We find both stable and unstable regions.  Consistent with the experimental observation of background gas collision limited lifetimes, we find that under the experimental conditions the condensate is stable against atomic collisions. Similar considerations will be important for any cold atom experiments on periodically driven systems. \\

The Schr\"{o}dinger equation with periodic driving is analyzed using Floquet theory \cite{Gomez-LeonFloquetDimensionPRL,Hanggireview}. Prior studies of periodically driven lattice systems have largely ignored interactions, focussing instead on how the single-particle physics is renormalized by the driving. For example, the band curvature and effective mass can be tuned with this technique \cite{EckardtFloquetPRL2005,LignierFloquetPRL2007,WernerBandFlippingPRL}. One can even invert a band, effectively flipping the sign of the hopping matrix elements. This latter feature has been used to realize models of frustrated magnets \cite{SengstockFrustratedScience2011,SengstockIsing2013}. More sophisticated driving techniques can be used for engineering artificial gauge fields \cite{SengstockAbelianGaugePRL2012,SengstockNonAbelianGaugePRL2012}. The driving can cause band-crossing leading to non-trivial topological numbers  \cite{GalitskiFTINatPhys2011,RechtsmanPFTI,ZhaiFTIarxiv,GaliskiFTIPRB2013,GedikFloquetScience,GalitskiFregosoFTIPRB2013,Gomez-LeonArxivAQHE}.  Extending these results to include interactions is important. Here, we look at atom-atom scattering. In the context of solid state physics, there has been some consideration of electron-phonon scattering \cite{DemlerFloquetTransport,DuttaSenPRB2013}. There also have been studies of non-dissipative interaction physics \cite{ChinZhaoarxiv}. Our work has connections with broader studies of heating in periodically driven systems \cite{PolkovnikovAnnals2013Floquet, RigolFloqetArxiv2014Floquet, EckardtPRLBoseSelection2013, DasMoessnerPRL,DasMoessner2014-2,SenguptaSensarma2013,ChandranAbaninMBLfloquet} . \\

In Section II, we describe the experiment and our main results about the stability of the condensate against atom-atom scattering. In section III A, we derive the Floquet spectrum and in Sec. III B, we predict the decay rate of a Floquet BEC.

\section{Model}
In Ref.\ \cite{ChinFloquet2013}, Parker et al. load a Bose-Einstein condensate (BEC) of 25,000 $^{133}{\rm Cs}$ atoms into a one-dimensional optical lattice. This lattice is then shaken at a frequency $\omega$, where  $\omega \approx (7.3 \times 2 \pi)$ kHz is slightly larger than $\frac{\Delta_0}{\hbar}$: $\Delta_0$ is the energy difference between the first and the second band at $k=0$. From the experimental parameters, we estimate $\Delta_0 \approx 4.96\,\ E_R$, where $E_R = \frac{h^2}{2 m \lambda_L^2}$ ($\lambda_L$ is the laser wavelength and is 1064 nm for this experiment).  The amplitude of shaking is slowly ramped up to a final value near 15-100 nm for a time of 5-100 ms. The shaking continues for 50-100 ms before the lattice and all the confinement is turned off, allowing the condensate to expand. By looking at the time of flight expansion images, the experimentalists determine if the condensate is at zero-momentum or finite momentum. By analogy with an Ising ferromagnet, where the condensate momentum is mapped onto the magnetization, they refer to these scenarios as paramagnetic and ferromagnetic. They also describe this as a $Z_2$ condensate.\\

In the frame of the moving lattice, the Hamiltonian for the driven system is given by $H=H_0(t) + H_{\rm int}$, where \cite{HolthausFloquetReview},
\bea
H_0(t) &=& \int\!dx\,\ \Psi^{\dagger}(x)\left(\frac{-\hbar^2}{2 m}\frac{d^2}{dx^2} + V_0 \sin^2\left(\frac{2 \pi x}{\lambda_L}\right) \right)\Psi(x) \nonumber\\
&+&\int\!dx \,\ \Psi^{\dagger}(x) \left(x F_0 \cos(\omega t) \right)\Psi(x), \\
H_{\rm int}&=& \frac{g}{2} \!\sum_{i1,i2,i3,i4} \int\!dx\,\ \Psi_{i1} ^{\dagger} (x)  \Psi_{i2} ^{\dagger}(x)  \Psi_{i3}  (x)  \Psi_{i4} (x).\nonumber\\
\eea
The atomic mass is m, the force from the periodic shaking is $F_0\cos(\omega t)$ and $g\approx \frac{4 \pi \hbar^2 a_s}{m d_{\perp}^2}$ is the 1-D effective interaction strength: $a_s$ is the scattering length and $d_{\perp}$ is the length-scale of transverse confinement.\\

The most intuitive way to analyze such a periodically driven system is to imagine observing the evolution of the system stroboscopically: i.e at times $t,t+T,t+2T,\ldots t+nT$; where $T=\frac{2 \pi}{\omega}$ is the time-period of the Hamiltonian and n is an integer. The time-evolution operator for n-periods is the n'th power of the time-evolution operator for one period:
\be
U(n T) = {\mathcal{T}}\exp \left( -i \int_{0} ^{T}\!dt \,\ H(t)/\hbar \right) = U(T)^n
\ee
It is therefore natural to define an effective Hamiltonian, $H_{\rm eff}$, such that
\be
U(T)= \exp(-i H_{\rm eff}T/\hbar)
\ee

In analogy to describing the labeling of Bloch states as ``quasi-momentum", the eigenvalues of $H_{\rm eff}$ are ``quasi-energies". The operator $H_{\rm eff}$ is not unique, as its eigenvalues (i.e ``quasi-energies") are only defined up to multiples of $\hbar \omega$. One can associate with each Bloch band of the undriven system, an infinite ladder of Floquet bands, separated by energies $\hbar \omega$. For the rest of the paper, we refer to the Bloch band connected adiabatically to the first (second) Bloch band in the limit of zero shaking as the ground (first excited) band.\\

Figure 1 shows typical Floquet bands for experiments analogous to Parker {\it et al.}'s. The ground band and the first excited band are shown by solid lines, their periodic repetitions by dashed lines. As is clear from the magnified views on the right, hybridization leads to a double well structure for the ground band. As illustrated by arrows, there are two classes of momentum and energy conserving scattering processes which can destabilize a BEC in one of the minima. These either involve scattering into two distinct bands (as in Fig.\ 1(a)) or into two periodic repetitions of the same band (Fig.\ 1(b)).  In Section III, we calculate the rate of scattering by Fermi's Golden rule \cite{SakuraiQM}. These scatterings are made possible due to the periodicity of energy for Floquet bands. \\
\begin{figure}
\includegraphics[scale=0.465]{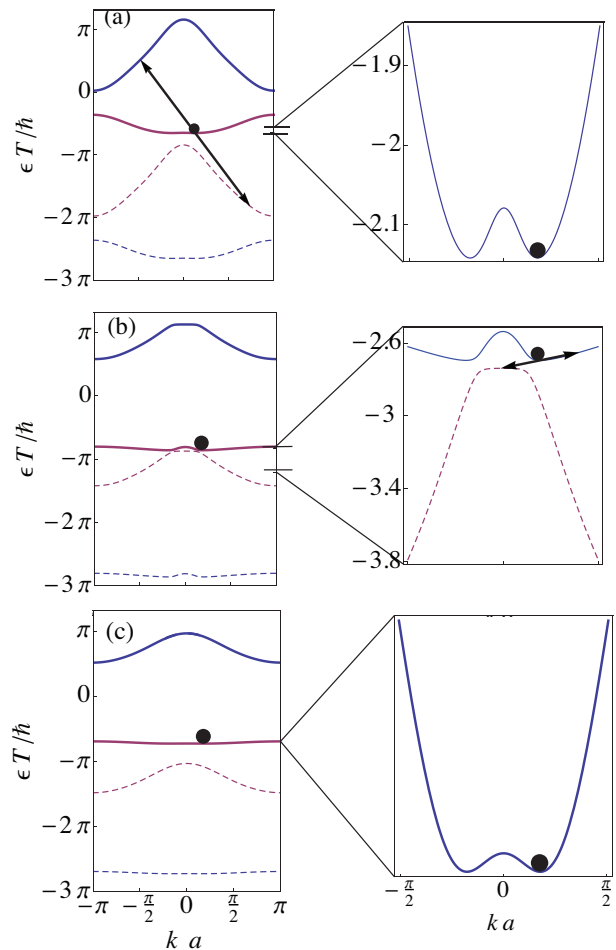}
\caption{(Color Online) Floquet Spectra of shaken 1 D lattices for lattice depths of $V_0/E_R =  2.02, 7$ and $7$. In (a) and (c), the shaking frequency is blue detuned while in (b), the shaking frequency is red detuned. The parameters in (c) are similar to those in Ref.\ \cite{ChinFloquet2013}. Quasi-momentum, $k$ and Quasi-energy, $\epsilon$ are measured in terms of the lattice spacing $a$ and the period, $T= \frac{2 \pi}{\omega}$. Solid and dashed lines represent bands and their periodic repetition and circles show location of band minima. Right panels are magnified views. In (a) and (b) arrows represent scattering processes which cause a condensate at the band minima to decay. In (a) this is an intra-band scattering process, where the final state of the scattered particles have the same Bloch index. In (b), it is an inter-band process where the Bloch indices are different. Case (c) is stable: there are no 2-body processes that conserve energy and momentum.}
\label{phase}
\end{figure}

As we explain in detail in Sec III, we use phase-space arguments to construct the phase diagrams in Fig.\ 2 and Fig.\ 3. As already introduced, we label phases as ferromagnetic or paramagnetic, depending on the momentum of the lowest energy Bloch state in the first band. In these diagrams, we also show if a condensate in that state is stable against 2-body collisions. Our model contains three relevant parameters: the detuning $(\hbar \omega - \Delta_0)$, the lattice depth $V_0$ and the shaking amplitude $F_0$. Fig.\ 2 shows a slice through the three dimensional phase diagram at $F_0 = 0$. The ferromagnetic regime (for infinitesimal shaking) is found when the two Floquet bands cross, i.e where $\Delta_{\pi} < \hbar \omega < \Delta_0$, where $\Delta_0$ is the band spacing at $k=0$ and $\Delta_{\pi}$ is the band spacing at $k = \pi$. This ferromagnetic phase is always unstable for infinitesimal $F_0$. Depending on parameters, the paramagnet may be stable or unstable.\\

Increasing the drive strength hybridizes the bands, generically increasing the ferromagnetic area. Somewhat counterintuitively the driving can stabilize or destabilize the system depending on parameters. This is a consequence of how the shaking modifies the bandstructure. Fig.\ 3 shows a representative slice at $V_0 = 7.0 E_R$, corresponding to the lattice used in Ref.\ \cite{ChinFloquet2013}. As $V_0$ is increased, both the unstable ferromagnetic phase and stable paramagnetic phase evolve into a stable ferromagnetic phase.\\

In principle, there may be kinematically allowed decay channels involving higher bands, but the rates will be very low due to small matrix elements. For very strong interactions, one should also include mean-field shifts to the bandstructure. These are irrelevant for Ref.\cite{ChinFloquet2013}, where the onsite interaction energy is $U_H =  0.001 E_R$ and the bandwidth $4 J = 0.16 E_R$. 
\begin{figure}
\begin{center}
\includegraphics[scale=0.55]{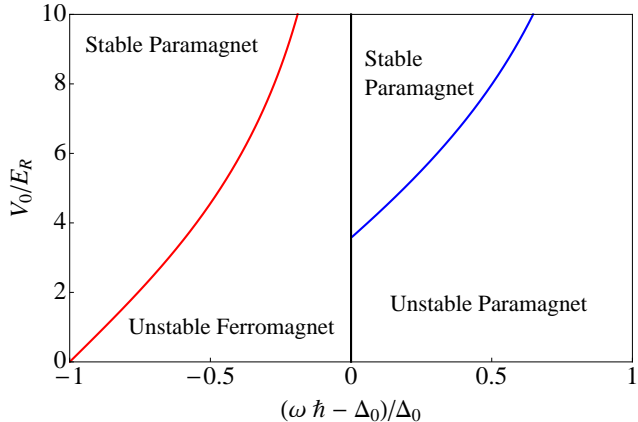}
\end{center}
\caption{(Color Online) Phase diagram of the floquet BEC for a variety of lattice depths and detunings in the limit of infinitesimal driving.}
\label{phase2}
\end{figure}

\begin{figure}
\begin{center}
\includegraphics[scale=0.55]{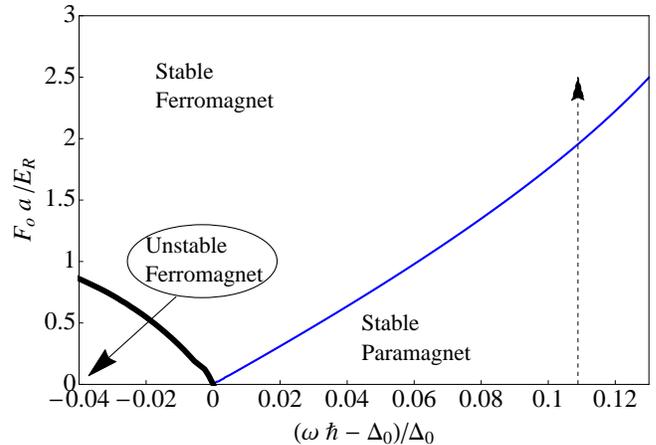}
\end{center}. 
\caption{(Color Online) Phase Diagram of the Floquet BEC in a shaken one-dimensional lattice of depth $V_0 = 7.0 \,\ E_R$. The zero-momentum bandgap, $\Delta_0$ is $4.96 \,\ E_R$. The vertical arrow shows the parameters of Ref.\cite{ChinFloquet2013}. The BEC is stable in the blue detuned regime. In the red detuned regime, the BEC is unstable below a critical driving strength and stable above it. The thick black line shows the critical driving strength.}
\label{phase1}
\end{figure}
\section{Calculation}
\subsection{Floquet Spectrum}
To derive the Floquet spectrum, we map the moving frame continuum Hamiltonian $H_0(t)$  onto a tight binding model. This is accomplished by expanding the field operator $\Psi(x)$ in terms of the Wannier functions for the two lowest bands of $H_0$ in the limit of vanishing $F_0$:
\be
\Psi(x) = \sum_j w_1(x-x_j)a_j +  w_2(x-x_j)b_j,
\ee
where $a_{j}$ and $b_j$ are bosonic annihilation operators and with the Wannier functions centered at the lattice site $n$ given by:
\be
w_{\sigma}(x-x_n) = \frac{1}{\sqrt{N}} \sum_k \exp(-i n k a) \psi_{\sigma} (x,k),
\ee 
where $N$ is the number of sites. The Bloch wave functions, $\psi_{\sigma} (x,k)$ are eigenstates of $H_0$ (with $F_0=0$) with 
\be
\psi_{\sigma} (x+a,k) = \exp(-i ka)  \psi_{\sigma} (x,k).
\label{blochw}
\ee
The arbitrary global phase of $\psi_{\sigma} (x+a,k)$ is fixed using the recipe given in Ref. \cite{zakblochband}.
The resulting tight-binding model is:
\be
H_0(t) = \sum_{ij} \left(-t_{ij} ^{(1)} a_{i}^{\dagger}a_{j} +  t_{ij} ^{(2)}  b_{i}^{\dagger}b_{j} + h.c.\right) +  \sum_{j} A_j(t),
\ee
where,
\bea
A_j(t) &=& F_0\cos(\omega t) \left(x_j \left(a_{j}^{\dagger} a_{j} + b_{j}^{\dagger} b_{j}\right) + \chi_j a_j^{\dagger} b_j + \chi_j ^{*} b_j^{\dagger} a_j \right)\nonumber  \\
\\
\chi_j  &=& \int\!dx \,\ x w_1^{*}(x-x_j) w_2(x-x_j) \nonumber \\
t_{ij} ^{(1)} &=& \int dx\,\ w_1^{*}(x-x_i)\left(\frac{-\hbar^2}{2 m} \frac{d^2}{dx^2}+ V(x)\right)w_1^{*}(x-x_j)\nonumber \\
t_{ij} ^{(2)} &=& \int dx\,\ w_2^{*}(x-x_i) \left(\frac{-\hbar^2}{2 m} \frac{d^2}{dx^2}+V(x)\right)w_2^{*}(x-x_j) \nonumber 
\eea
with $V(x) = V_0 \sin^2\left(\frac{2 \pi x}{\lambda_L}\right)$. Equivalently, we find $t_{ij}^{\sigma}$ by fitting the dispersion obtained from the tight-binding model to the dispersion of the Bloch bands. For the experimental lattice strength, the ground band is well approximated by a model with nearest neighbor hopping. However, to properly account for the greater curvature of the first excited band, one needs to take into account longer range hopping (up to $\vert i-j\vert \le 3$).\\

We now rotate our basis, taking $\vert\psi\rangle \rightarrow U_{c}(t) \vert \psi\rangle$ with:
\be
U_{c}(t) = \exp\left(- \frac{i}{\hbar} \int_{0}^{t} \sum_{j}x_ j F_0\cos(\omega t) (a_{j}^{\dagger} a_{j} + b_{j}^{\dagger} b_{j}) \right)
\label{unitary}
\ee
Under this unitary transformation, the Hamiltonian becomes:
\bea
H_0^{\prime} (t) &=& U_{c}H_0(t)U_{c}^{-1} - i \hbar U_{c}\partial_t U_{c}^{-1} \nonumber  \\
&=& \sum_{ij} \left(-J_{ij} ^{(1)} (t) a_{i}^{\dagger}a_{j} +  J_{ij} ^{(2)}(t)  b_{i}^{\dagger}b_{j} + h.c.\right) \nonumber\\
&+&\sum_j F_0\cos(\omega t)  \left(\chi_j a_{j}^{\dagger} b_{j} +\chi_j ^{*} b_{j}^{\dagger} a_{j}  \right) \nonumber \\
&=& \sum_k \sum_m  \cos(m k a)\left(-J_{m} ^{(1)} (t) a_{k}^{\dagger}a_{k} -J_{m} ^{(2)} (t) b_{k}^{\dagger}b_{k}\right)\nonumber \\
&+& \sum_k F_0\cos(\omega t)  \left(\chi_j a_{k}^{\dagger} b_{k} +\chi_j ^{*} b_{k}^{\dagger} a_{k} \right) \nonumber \\
\eea
where,
\bea
J_{ij}^{\sigma} (t) &=& t_{ij}^{\sigma} \exp(-i F_0 \frac{\cos(\omega t)}{\hbar \omega} (x_i-x_j)) \nonumber\\
&=& t_{ij}^{\sigma} \exp(-i F_0 \frac{\cos(\omega t)}{\hbar \omega} a (i-j)) ,
\label{rwa1}
\eea
the lattice spacing is a and $m=\vert i-j\vert = \{1,2,3\}$. We numerically calculate the time evolution operator, $U(T)$ by integrating $i \hbar \partial_t U = H U$ from $t=0$ to $t=T = \frac{2 \pi}{\omega}$ with the boundary condition $U(0) = \mathds{1}$. The ``quasi-energies", $\epsilon$ are given by the eigenvalues of the matrix $(i \hbar/T)  \log[U(T)]$. We stress again that since, a logarithm has an infinite number of branches, the energy spectrum is unbounded. Typical results are shown in Fig.\ 1.\\

\subsection{Rotating Wave Approximation}
While the scattering rate may be calculated by the Floquet formalism, we can simplify the argument by making a Rotating Wave Approximation which is the leading order expansion in $F_0 a/ \hbar \omega$. We will calculate the rates in the region where $\frac{F_0 a}{\hbar \omega} \approx 0.005$. In this limit, Eq.(\ref{rwa1}) reduces to $J_{ij}^{\sigma} (t) = t_{ij}^{\sigma}$ is time-independent. Thus,we obtain an effective Hamiltonian :
\bea
H_{\rm eff} (k) &=& \sum_k  \left(E^{(1)}_k  a_{k}^{\dagger}a_{k} + E^{(2)}_k  b_{k}^{\dagger}b_{k}\right)\nonumber \\
&+& \sum_k F_0 \left(\exp(- i \omega t)  \chi_j a_{k}^{\dagger} b_{k} + \exp(i \omega t) \chi_j ^{*} b_{k}^{\dagger} a_{k} \right)\nonumber ,\\
\eea
where
\bea
E^{(1)}_k &=& -\sum_m  \cos(m k a) t_{m} ^{0}\,\,\ {\rm and} \nonumber \\
E^{(2)}_k &=& -\sum_m  \cos(m k a) t_{m} ^{(1)} \nonumber ,
\eea
where $t_{m} ^{\sigma} = t_{i,i+m} ^{\sigma}$ and is the same for any site $i$ since the system is homogenous. Further, under the canonical transformation, $U=\exp\left(i \omega t \,\ b_k ^{\dagger} b_k\right)$, the Hamiltonian takes the form:
\bea
H_{\rm eff} (k) &=& \sum_k  \left(E^{(1)}_k  a_{k}^{\dagger}a_{k} + (E^{(2)}_k - \hbar \omega)   b_{k}^{\dagger}b_{k}\right)\nonumber \\
&+& \sum_k F_0 \left(  \chi_j a_{k}^{\dagger} b_{k} + \chi_j ^{*} b_{k}^{\dagger} a_{k} \right)\nonumber \\
\label{eff}
\eea

We use this Hamiltonian for calculating the scattering rate using Fermi's golden rule.
\subsection{Scattering Rate}
Since Eq.(\ref{eff}) is time-independent, we can use Fermi's golden rule \cite{SakuraiQM} to calculate the rate for two particles to scatter from initial state $\vert \psi_i\rangle$ to final state $\vert \psi_f\rangle$ as: 
\be
\frac{d N}{dt} = \frac{2 \pi}{\hbar} \sum_n \vert \langle \psi_f \vert H_{\rm int} \vert \psi_i \rangle \vert^2  \delta(E_f-E_i),
\ee
For our calculation, $\vert \psi_i \rangle$ corresponds to the BEC at momentum $k_0$, while $\vert \psi_f\rangle$ has two particles outside of the condensate:
 \be
 \vert \psi_i\rangle =\frac{ (\Phi_{1} ^{\dagger} (k_0))^N}{\sqrt{N!}} \vert {\rm vac} \rangle \nonumber
 \ee
and
\be
 \vert \psi_f\rangle =\frac{(\Phi_{i_1} ^{\dagger} (k_0 + q))(\Phi_{i_2} ^{\dagger} (k_0-q)) (\Phi_{1} ^{\dagger} (k_0))^{N-2}}{\sqrt{(N-2)!}} \vert {\rm vac} \rangle \nonumber
 \label{fgr}
 \ee
where $\Phi_i ^{\dagger} (k)$ is the boson creator operator at momentum $k$ is the dressed band $i$. Kitagawa {\it et al.} \cite{DemlerFloquetTransport} generalize Eq.(\ref{fgr}) to the situation where the rotating wave approximation breaks down. \\ 
 
 We expand the field operator in terms of the Bloch functions in Eq. (\ref{blochw}),
\be
\Psi_{\sigma} (x) = \sum_k \overline{\Phi_{\sigma}} (k) \psi_{\sigma} (x,k).
\ee
This yields an interaction Hamiltonian of the form
\bea
\frac{H_{\rm int}}{g} &=& \frac{1}{2}\left(\sum_{j} \int_{0} ^{L} dx\,\  \Psi_{j} ^{\dagger} (x)  \Psi_{j} ^{\dagger}(x)  \Psi_{j}  (x)  \Psi_{j} (x)\right) \nonumber \\
&+&2\left( \int_{0} ^{L} \! dx\,\   \Psi_{1} ^{\dagger} (x)  \Psi_{2} ^{\dagger}(x)  \Psi_{1}  (x)  \Psi_{2}  (x) \right) \nonumber \\
&=&\frac{1}{2} \left(\sum_{\{k\}, j}  \Gamma_{j\,\ j\,\ j\,\ j} ^{k_1k_2k_3k_4} \overline{\Phi_{j}} ^{\dagger}  (k_1)  \overline{\Phi_{j}}^ {\dagger} (k_2)  \overline{\Phi_{j}} (k_3)  \overline{\Phi_{j}} (k_4) \right)\nonumber\\
&+& 2\left(\sum_{\{k\}}  \Gamma_{1\,\ 2\,\ 1\,\  2} ^{k_1k_2k_3k_4} \overline{\Phi_{1}} ^{\dagger}  (k_1)  \overline{\Phi_{2}}^ {\dagger} (k_2)  \overline{\Phi_{1}} (k_3)  \overline{\Phi_{2}} (k_4) \right)  \nonumber.\\
\label{intham}
\eea
where, the index $j$ labels the Bloch band and $\{k\} = \{k_1, k_2, k_3, k_4\}$. The matrix elements are :
\be
\Gamma_{i_1i_2i_3i_4} ^{k_1k_2k_3k_4} =  \int_{0} ^{L}\! dx\,\ \psi_{i_1} ^{*} (x,k_1)  \psi_{i_2} ^{*}(x,k_2)  \psi_{i_3}  (x,k_3)  \psi_{i_4} (x,k_4)\nonumber. \\
\ee
\\
$\Gamma_{i_1i_2i_3i_4} ^{k_1k_2k_3k_4}$ vanishes unless $k_1+k_2 = k_3+k_4 + 2\pi m/a$ for some integer $m$. Due to this phase space constraint, there are values of $\omega$ and $F_0$ for which no scattering is possible. Fig 2 shows the stability phase diagram for $V_0 =  7 E_R$, where $\Delta_0 = 4.96 E_R$. \\

\begin{figure}
\begin{center}
\includegraphics[scale=0.57]{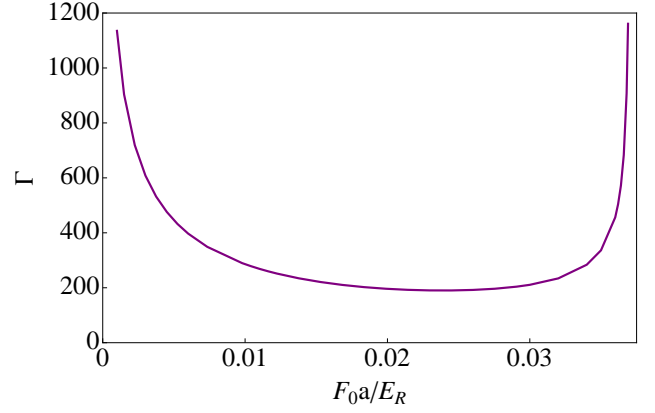}
\end{center}
\caption{(Color Online) Plot of dimensionless decay rate $\Gamma$ (defined in Eq.(\ref{gammadef}) in the text) as a function of amplitude of shaking, $F_0 a/E_R$ for $\omega = 4.95 \,\ E_R/\hbar$ and $V_0 = 7.0 E_R$. Within the rotating wave approximation, with this $\omega$ the transition from unstable ferromagnetic phase to stable ferromagnetic phase happens at $F_0 = 0.037 \,\ E_R/a$.}
\label{phase3}
\end{figure}


In the unstable region, the matrix element in Fermi's Golden rule takes the form:
\bea
\vert \langle \psi_f \vert H_{\rm int} \vert \psi_i \rangle \vert^2 &=&  N(N-1) \frac{g^2}{4} \sum_q \vert \sum_{i1 i2} F_{i_1i_2} ^{k_0+qk_0-qk_0k_0}\vert^2 \nonumber \\
&\approx&  N^2 \frac{U^2}{4} \sum_q \vert \sum_{i1 i2} F_{i_1i_2} ^{q k_0}\vert^2 \nonumber\\
&=& N^2 \frac{U^2}{4} \frac{L}{2\pi} \int dq  I^{q k_0}
\label{fgr}
\eea
where,
\be
I^{q k_0}=  \vert \sum_{i1 i2} F_{i_1i_2} ^{q k_0}\vert^2 \nonumber
\ee

Hence, we see that the scattering rate is given by:
\bea
\frac{d N}{dt} &=& \sum_{n j_1 j_2} \frac{2 \pi}{\hbar}\frac{L}{2\pi}N^2 \frac{g^2}{4} \int\! dq I^{q k_0}  \delta(E_f-E_i)\nonumber\\
&=& \sum_{n j_1 j_2} \frac{N^2 g^2 L}{4 \hbar}  \int\! dE_f \frac{dq}{dE_f} I ^{q k_0}  \delta(E_f-E_i )\nonumber\\
&=&\sum_{n j_1 j_2} \frac{g^2}{4 \hbar E_R}\frac{N^2}{La} L^2 \int\! dE_f \frac{d(qa)}{d (E_f/E_R)} I ^{q k_0}  \delta(E_f-E_i) \nonumber\\
\label{gammadef}
&=&  \frac{g^2}{4 \hbar E_R}  \frac{N^2}{La} \Gamma,
\eea
which defines the intensive dimensionless quantity $\Gamma$, which depends only on the lattice geometry and the shaking strength. Typical behavior is shown in Fig.\ 4. At the threshold for dissipation, the scattering rate diverges. This is a consequence of the 1-D density of states. Away from these singularities, $\Gamma \approx 200$ for this lattice depth and shaking frequency. Taking $n= N/L \approx 1/a$, $a_s \approx 2$ nm and $d_{\perp} \approx  100 $ nm yields $\tau = N/(dN/dt) \sim 0.06$ ms.

\section*{Acknowledgements}
This paper is based on work supported by the National Science Foundation under Grant no. PHY-1068165 and from ARO-MURI Non-equilibrium Many-body Dynamics grant (63834-PH-MUR).

\end{document}